\documentclass[nonatbib]{article}

\usepackage[final]{neurips_2022}
\usepackage[colorlinks=true,citecolor=blue]{hyperref}
\usepackage{physics}
\usepackage{verbatim}
\usepackage[backend=biber,sorting=none,citestyle=numeric-comp]{biblatex}
\usepackage[utf8]{inputenc} 
\usepackage[T1]{fontenc}    
\usepackage{hyperref}       
\usepackage{url}            
\usepackage{booktabs}       
\usepackage{amsfonts}       
\usepackage{amsmath,amssymb}
\usepackage{nicefrac}       
\usepackage{microtype}      
\usepackage{xcolor}         

\usepackage{graphicx, subcaption}
\usepackage{dcolumn}
\usepackage{bm}
\usepackage{xcolor}
\usepackage{multirow}
\usepackage{caption}
\usepackage{float}
\DeclareRobustCommand{\Sec}[1]{Sec.~\ref{#1}}

\DeclareRobustCommand{\Tab}[1]{Table~\ref{#1}}

\DeclareRobustCommand{\Fig}[1]{Fig.~\ref{#1}}

\title{Efficiently Moving Instead of Reweighting Collider Events with Machine Learning}

\author{Radha Mastandrea \\
Department of Physics, University of California, Berkeley \\
Berkeley, CA 94720 \\
Physics Division, Lawrence Berkeley National Laboratory, Berkeley \\
Berkeley, CA 94720 \\
\texttt{rmastand@berkeley.edu}
\And
  Benjamin Nachman \\
Department of Physics, University of California, Berkeley \\
Berkeley, CA 94720 \\
Physics Division, Lawrence Berkeley National Laboratory, Berkeley \\
Berkeley, CA 94720 \\
Berkeley Institute for Data Science, University of California, Berkeley \\
Berkeley, CA 94720 \\
\texttt{bpnachman@lbl.gov}
}

\begin{document}

\maketitle


\begin{abstract}
There are many cases in collider physics and elsewhere where a calibration dataset is used to predict the known physics and / or noise of a target region of phase space.  This calibration dataset usually cannot be used out-of-the-box but must be tweaked, often with conditional importance weights, to be maximally realistic.  Using resonant anomaly detection as an example, we compare a number of alternative approaches based on transporting events with normalizing flows instead of reweighting them.  We find that the accuracy of the morphed calibration dataset depends on the degree to which the transport task is set up to carry out optimal transport, which motivates future research into this area.

\end{abstract}

\section{Introduction}
\label{sec:introduction}

Many tasks in collider physics depend on the calibration of auxiliary datasets.  Data from a well-understood region of phase space are chosen as a \textit{reference} to model the known physics in a \textit{target} region of phase space.  When the reference and known physics in the target are identically distributed, differences between the reference and target would indicate the presence of new phenomena.  However, the reference data may be a distorted version of the known physics in the target and so corrections need to be applied. The reference is chosen to be as similar as possible to the known physics in the target so that the corrections applied should be small.

Traditionally, this calibration task has been performed using importance weights estimated from ratios of histograms, either using data-driven approaches like the control region method or fully data-based approaches like the ABCD or Matrix Method techniques (see e.g. Ref.~\cite{Karagiorgi:2021ngt} for a review).  These methods can be generalized to the unbinned (and high-dimensional) case with machine learning-based likelihood ratio estimation~\cite{Andreassen:2020nkr,ATLAS:2022hwc,ATLAS:2022hra}.  An alternative approach that may be more precise, especially when the reference and target probability densities have regions of non-overlapping support, is to morph the features themselves.  Given a reference distribution $X_R\sim p_{X_R}$ and a target distribution $X_T\sim p_{X_T}$, a mapping function $f:X_R\rightarrow X_T$ is chosen so that the probability density of the transformed reference $f(X_R)$ is as close a match as possible to the probability density of the target.

Optimal transport (OT) has been studied in collider physics to solve the calibration problem~\cite{Pollard:2021fqv}.  In this paper, we focus on the case of conditional morphing, where $X_R$ and $X_T$ are conditioned on an observable $M$ (often a mass, thus the symbol), and we set $f(\cdot|M):X_R|M\rightarrow X_T|M$.  The reason for this setup is that the morphing function is typically learned in a background-dominated region and then interpolated or extrapolated in $M$ to a signal-sensitive region.  This setup was first explored in collider physics in Ref.~\cite{Raine:2022hht} and uses normalizing flows (NFs)~\cite{https://doi.org/10.48550/arxiv.1505.05770}.  These NFs are not specifically tasked with making the morphing minimal. This our goal in this paper is to explore how minimal these moves are and examine if tweaks to the setup can bring these moves closer to the OT solution and thus improve model fidelity for downstream inference tasks.  As in Ref.~\cite{Raine:2022hht}, we use anomaly detection as our numerical example.


This paper is organized as follows. In \Sec{sec:methods}, we introduce the application of normalizing flows to the problem of resonant anomaly detection and describe our dataset and training procedure.  Numerical results are presented in \Sec{sec:results}, and we discuss future directions in Ref.~\Sec{sec:future_work}.

\section{\label{sec:methods}Methods}

\textit{Resonant} or \textit{group} anomaly detection (see e.g. Ref.~\cite{Kasieczka:2021tew}) in collider physics begins with a resonant feature $M$.  A potential signal will have $|M-M_0| \lesssim c$ (which defines the signal region) for some unknown $M_0$ and often knowable $c$.  The value of $M_0$ is usually found through a scan.  Additional features $X\in\mathbb{R}^N$ are chosen which can be used to distinguish signal from background.  Weakly supervised methods learn a classifier acting on $X$ that can distinguish signal region events in the target from events in the reference~\cite{Collins:2018epr,Collins:2019jip,Nachman:2020lpy,Andreassen:2020nkr,1815227,Hallin:2021wme,Raine:2022hht}.  Approaches vary on how they construct the reference, but most use sideband information ($|M-M_0|\gtrsim c$) to some degree.  The first use of NFs in this context was in Ref.~\cite{Nachman:2020lpy}, which used NFs as a generative model.  In contrast, the idea of Ref.~\cite{Raine:2022hht} is to use NFs as the morphing functions described in the previous section.



Normalizing flows are neural networks that can approximate the density of complex data through a composition of relatively simpler functions with a tractable Jacobian: $L=\log p(z)+\sum \log J_i$, where $L$ is the loss, $p(z)$ is the base distribution, and $J_i$ is the Jacobian when transforming from the $i^\text{th}$ function to the $(i+1)^\text{th}$ function in the composition.  We consider multiple ways of training the flow:


\begin{enumerate}
    \item \textbf{Double Base}.  We learn two conditional NFs, one that maps from a standard normal $z\sim\mathcal N(0,1)^N$ to the target $f_T(\cdot|M):z|M\rightarrow X_T|M$ and one from a standard normal to the reference $f_R(\cdot|M):z|M\rightarrow X_R|M$.  The morphing function is then $f_T\circ f_R^{-1}$.
    \item \textbf{Base to Data}.  We learn a conditional NF for the reference $f_R(\cdot|M):z|M\rightarrow X_R|M$ and then use this as the base density to learn a second flow from the reference to the target $f(\cdot|M):X_R|M\rightarrow X_T|M$.
    \item \textbf{Identity Initialization.} Same as (2), but we initialize the mapping at the identity function. This is done by first learning the mapping $f(\cdot|M):X_R|M\rightarrow X_R|M$, then using transfer learning to adapt this flow to map $X_R$ to $X_T$.
    \item \textbf{Movement penalty.} Same as (2), but we add a term to the loss function that explicitly penalizes movement in the parameter space, $L_2 = \alpha\lVert X_R - f( X_{R} |M) \rVert^2$ for a hyperparameter $\alpha$. 
    
\end{enumerate}

To evaluate the performance of the models, we study how far the events were moved, $|X_R - f( X_{R} |M) |$ as well as the fidelity of the move, quantified by the (ideally poor) performance of a post-hoc classifier trained to distinguish the interpolated reference from the target in the signal region.

\subsection{Dataset}
\label{sec:dataset}

\begin{figure}[h!]
    \centering
    \includegraphics[width = \linewidth]{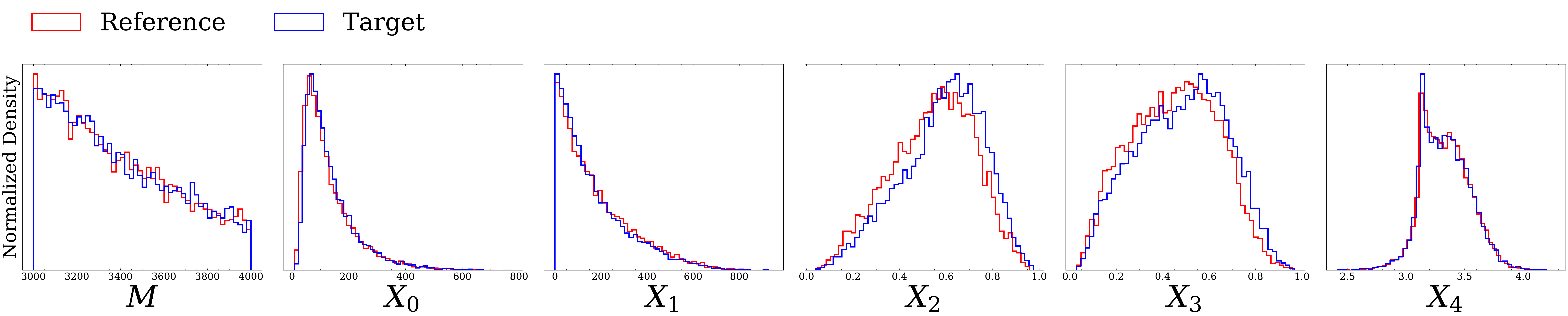}
    \caption{Reference and target distributions used in this study The feature space is comprised of the resonant feature $M$ and five other features $m_{J_1}$, $\Delta m_{JJ}$, $\tau^{21}_{J_1}$, $\tau^{21}_{J_2}$, and $\Delta R_{JJ}$. A description of these observables can be found in \cite{Thaler:2010tr}. The signal region is defined by $|M-M_0|<c$ for $M_0=3500$ and $c=100$.}
    \label{fig:LHC_datasets}
\end{figure}

We use the LHC 2020 Olympics R\&D dataset~\cite{LHCOlympics,Kasieczka:2021xcg} which consists of 1M background (Standard Model) events and 100k signal/anomaly events.  Our setup is nearly the same as in many previous resonant anomaly detection studies with the same dataset~\cite{Collins:2018epr,Collins:2019jip,Nachman:2020lpy,Andreassen:2020nkr,1815227,Hallin:2021wme,Raine:2022hht}.  The events naturally live in a high-dimensional space (each containing hundreds of particles with a momentum), but we consider a five-dimensional compressed version that has been extensively studied in collider analyses.  These five features, along with the resonant feature, are displayed in Fig.~\ref{fig:LHC_datasets}. The reference dataset consists of 1M events from one simulator while the target consists of 1M background events from a different simulator.  As we are investigating the background estimation, no anomalous events are part of the training or testing in this paper.

\subsection{\label{sec:training}Training procedure}

To generate flows that map from $z\sim\mathcal{N}(0,1)^N$ to a reference / target distribution, we use the highly expressive MADE autoregressive (AR) module \cite{https://doi.org/10.48550/arxiv.1502.03509}. We use 8 modules consisting of 8 stacked piecewise rational quadratic transformations, interleaved with a reverse permutation of all dimensions. Each masked linear layer of the MADE modules has dimensions (64, 64). The resonant feature $M$ is embedded in a linear network of size (1, 64). We train for 60 epochs with a batch size of 128, a a cosine annealing learning rate initialized at $10^{-4}$, and a weight decay of $10^{-4}$. 

To generate flows that map between the reference and target distributions (which are relatively similar to each other), we use a more lightweight architecture consisting of a coupling model with piecewise rational quadratic transformations. These flows contains 2 stacked layers interleaved with a reverse permutation, with each masked linear layer having dimensions (16, 16). We train for 40 epochs with a batch size of 256, a cosine annealing learning rate initialized at $4\times10^{-4}$, and a weight decay of $10^{-4}$. 

The training dataset is comprised from the sidebands of the resonant feature $M$, which ensures that the signal region is kept blinded until testing. All flows are constructed using the \textsc{nflows} package \cite{nflows}, trained using \textsc{PyTorch} \cite{NEURIPS2019_9015}, and optimized using \textsc{Adam} \cite{adam}. All hyperparameters are optimized through grid search. For each training method, we repeat the flow training 5 times with a different random seed.

\section{\label{sec:results}Results}

We evaluate the learned mappings from reference $X_R$ to target $X_T$ for the four training procedures proposed in \Sec{sec:methods} in two ways: (1) we observe how far in parameter space points from the reference travel as they are mapped to the target distribution, and (2) we evaluate the fidelity of the mapping through training a classifier to discriminate $X_T$ from $f( X_{R} |M)$. 

The mappings are evaluated in the signal region ($|M - M_0| < 100$), in training sidebands regions ($100 < |M - M_0| < 300$), and additionally in outer sidebands regions ($300 < |M - M_0| < 500$). This allows us to gauge the performance of the mapping in both interpolation and extrapolation tasks.

\subsection{Distanced moved in parameter space}

\begin{figure}[h!]
    \centering
    \includegraphics[width = \linewidth]{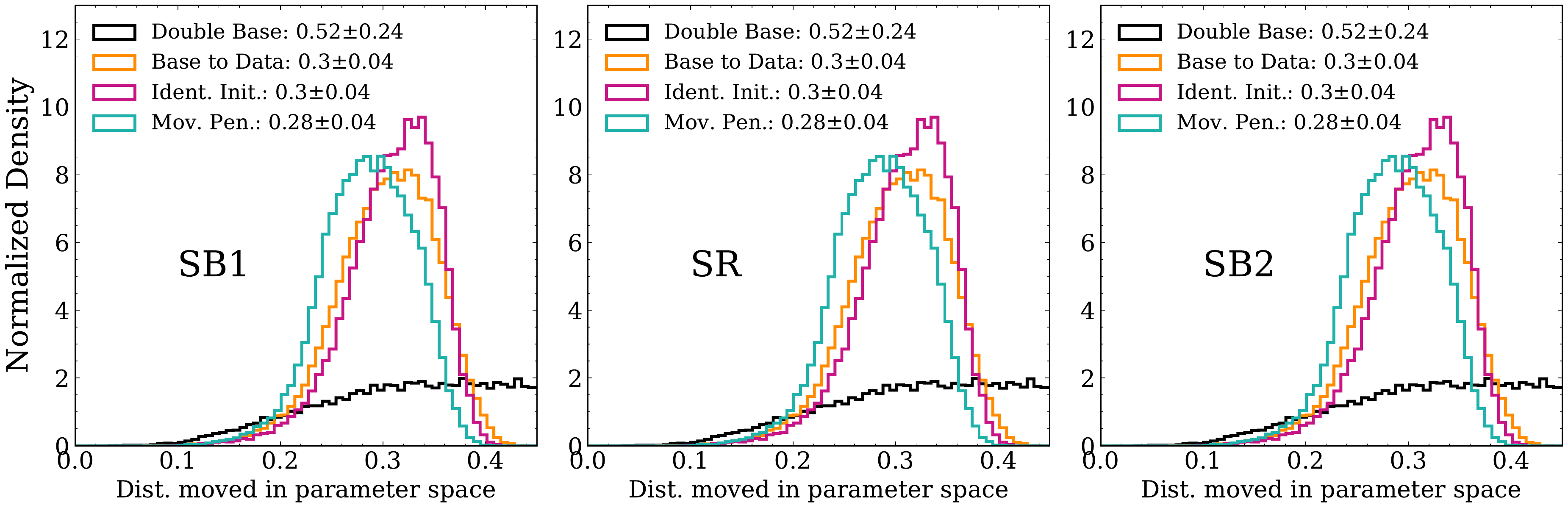}
    \caption{Distance traveled in the full parameter space under a mapping from reference $X_R$ to target $X_T$.  The data is preprocessed so feature lies in the interval (-3, 3), so the maximum possible distance traveled is $\sim13$. We provide the mean distances traveled $\pm 1\sigma$ errors.}
    \label{fig:distances}
\end{figure}

In \Fig{fig:distances}, we show the distances traveled in feature space for each data point $X_R$ that is mapped to the target distribution by $f( X_R| M)$. In \Sec{sec:app}, we show the distances traveled for features $X_2$ and $X_3$, which according to \Fig{fig:LHC_datasets} are the ones that are most different between the reference and target distributions. 

The Double Base method produces a vastly larger range of distances traveled than the other three methods, which is to be expected as the training procedure contains no explicit link between the reference and target distributions. The Movement Penalty method has the smallest mean distance traveled in the entire and individual feature space(s), although by a narrow margin.

\subsection{\label{sec:fot}Fidelity of transform}

In \Tab{tab:rocs}, we provide the ROC scores for a binary classification neural net trained to discriminate $f( X_R| M)$ from $X_T$. The classifiers are dense networks of three hidden layers and 32 nodes, all with $\textsc{ReLu}$ activation. We train the binary classifiers for 20 epochs with a batch size of 128 and a cosine annealing learning rate initialized at $10^{-3}$,

For the purpose of this report, we define a ``random" classifier to have a ROC score $< 0.51$. A successful mapping between the reference and target distributions would result in the trained binary classifier performing no better than random.

\begin{table}[]
    \centering
    \begin{tabular}{||c|c|c|c|c||}
    \hline
    \hline
    Band & Double Base & Base to Data & Identity Init. & $L_2$ ($\alpha$ = $10^{-2}$)  \\
    \hline
    \hline
    OB1 & 0.630 $\pm$ 0.024 & 0.511 $\pm$ 0.003 & 0.508 $\pm$ 0.004 & 0.507 $\pm$ 0.002 \\ 
    \hline
    SB1 & 0.501 $\pm$ 0.000 & 0.502 $\pm$ 0.001 & 0.501 $\pm$ 0.000 & 0.502 $\pm$ 0.001 \\ 
    \hline
    SR & 0.553 $\pm$ 0.011 & 0.503 $\pm$ 0.001 & 0.503 $\pm$ 0.001 & 0.503 $\pm$ 0.000 \\ 
     \hline
    SB2  & 0.501 $\pm$ 0.000 & 0.503 $\pm$ 0.001 & 0.503 $\pm$ 0.001 & 0.502 $\pm$ 0.001 \\ 
    \hline
    OB2 & 0.594 $\pm$ 0.030 & 0.506 $\pm$ 0.002 & 0.507 $\pm$ 0.004 & 0.507 $\pm$ 0.003 \\ 
    \hline
    \hline 
    \end{tabular}
    \\
    \caption{ROC scores for a binary classifier trained to discriminate mapped reference $f( X_R| M)$ from target $X_T$ . The scores are calculated separately for different regions of the resonant parameter. (OB1, OB2) = (low, high) mass outer bands; (SB1, SB2) = (low, high) mass sidebands; SR = signal region. We provide the mean ROC scores $\pm 1\sigma$ errors.}
    \label{tab:rocs}
\end{table}

All four training procedures show good fidelity in the sidebands regions (recall that this comprises the training dataset region). All methods except the Double Base method also show good fidelity when evaluated in the signal region. All methods exhibit a drop in performance when evaluated in the outer bands regions, but only the Double Base method drops enough to lose fidelity.

\section{\label{sec:future_work}Future areas of study}

In this work, we have explored modifications to the training procedure for a normalizing flow tasked with learning a mapping between a reference and a target dataset. Our explored application uses reference and target datasets that are similar, as might be expected in the collider physics calibration problem of modeling known physics in a specific region of phase space.

The Double Base method is clearly suboptimal in terms of optimal transport and mapping fidelity, likely due to the fact that it transports between the reference and the target through an uncorrelated standard normal distribution. The other three methods (Base to Data, Identity Initialization, and Movement Penalty) show comparable performances in both metrics, and in fact all meet the requirement of being a faithful mapping due to the indiscriminability of $X_T$ from $f(X_R|M)$. In this situation where the reference and target sets are similar, using the Base to Data method is acceptable. However, this method may not be adequate for more complex (transformations between) datasets.

As an avenue for future investigations, we may consider exchanging a normalizing flow for a continuous normalizing flow (CNF) \cite{https://doi.org/10.48550/arxiv.1806.07366}. Such a flow  restricts transformations to be continuous, such that each point can be assigned a trajectory with a velocity vector. Building upon this, the OT-Flow method \cite{DBLP:journals/corr/abs-2006-00104} adds to the CNF loss both an $L_2$ movement penalty and a penalty that encourages the mapping to transport points along the minimum of some potential function. Such alternatives might be explored for situations when the reference and target distributions are significantly different.

\section{Potential Broader Impacts}

In this work, we have explored methods to augment the training of normalizing flows that learn mappings between probability distributions derived from LHC-like particle collision datasets. However, these modifications could be used to improve the performance for any normalizing flow-like architecture, regardless of the physical origin of the reference and target datasets. Our work could then be beneficial to any physical problem that relies on the creation or transformations of probability distributions. Since our work serves as a method to improve an existing architecture, rather than to define a new analysis procedure, it is our belief that this work does not present any foreseeable societal consequence.

\section*{Code availability}

\noindent The code for all experiments in this report can be found at \href{https://github.com/rmastand/FETA}{https://github.com/rmastand/FETA}.

\begin{ack}

We thank Samuel Klein for his useful discussions relating to these experiments. B.N. and R.M. were supported by the Department of Energy, Office of Science under contract number DE-AC02-05CH11231. This material is based upon work supported by the National Science Foundation Graduate Research Fellowship Program under Grant No. DGE 2146752. Any opinions, findings, and conclusions
or recommendations expressed in this material are those of the authors and do not necessarily reflect the views of the National Science Foundation.

\end{ack}

\printbibliography



\clearpage
\appendix

\section{\label{sec:app}Appendix}

Here, we provide supplementary plots for the distances traveled in $X_2$ and $X_3$ space for each data point $X_R$ that is mapped to the target distribution by $f( X_R| M)$. The other three features did not show appreciable movement after training, likely due to the similarity of their distributions between the reference and the target datasets.

\begin{figure}[h]
     \centering
     \begin{subfigure}[b]{\textwidth}
         \centering
         \includegraphics[width=\textwidth]{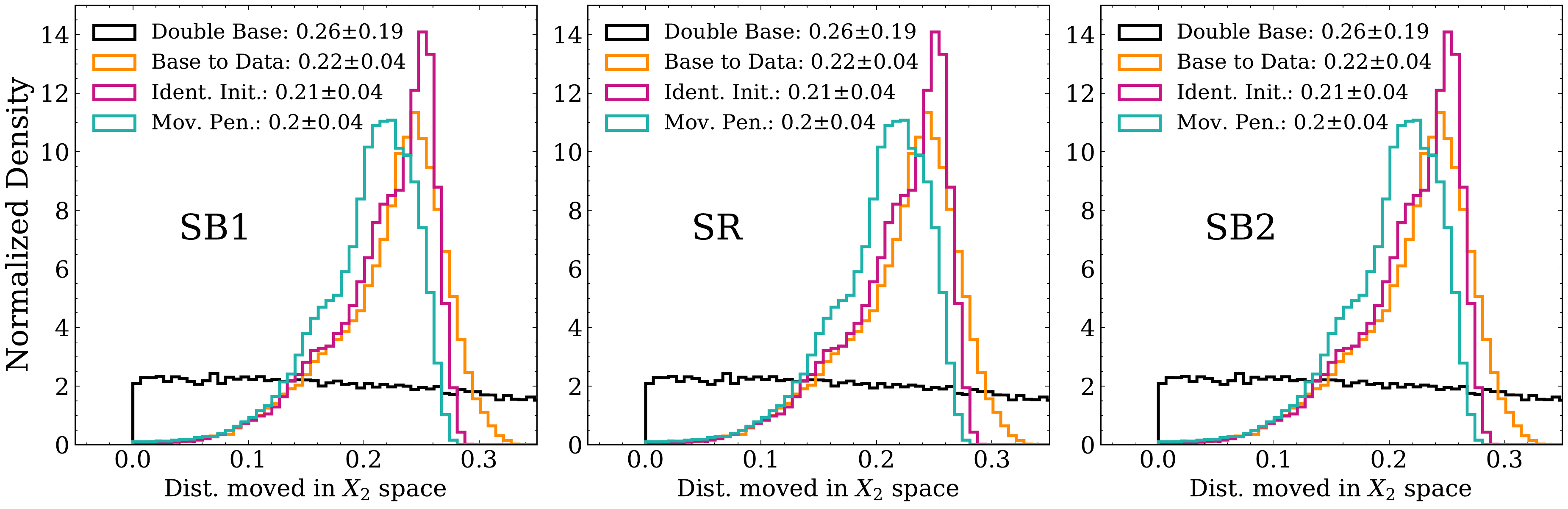}
         \caption{Feature $X_2$, corresponding to $\tau^{21}_{J_1}$. }
         \label{fig:distances_f2}
     \end{subfigure}
     \hfill
     \begin{subfigure}[b]{\textwidth}
         \centering
         \includegraphics[width=\textwidth]{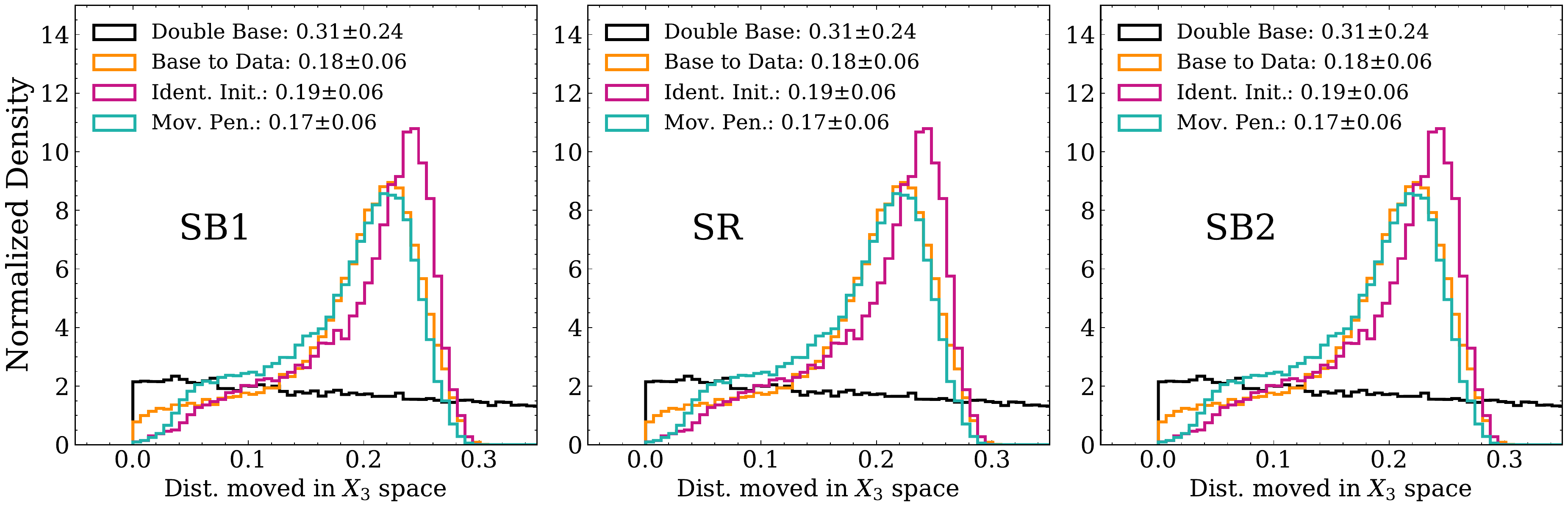}
         \caption{Feature $X_3$, corresponding to $\tau^{21}_{J_2}$.}
         \label{fig:distances_f2}
     \end{subfigure}
  
        \caption{ Distance traveled in a single dimension of the parameter space under a mapping from reference $X_R$ to target $X_T$. The maximum possible distance traveled is 6. We provide the mean distances traveled $\pm 1\sigma$ errors.}
        \label{fig:distances_feature}
\end{figure}

\end{document}